\documentclass[pra,aps,twocolumn,floatfix,showpacs]{revtex4}
\usepackage{graphicx,graphics,psfrag,epsfig}
\begin{document}

\title{The evolution from BCS to BEC superfluidity
in the presence of disorder}

\author{Li Han and C. A. R. S{\'a} de Melo}

\address{School of Physics, Georgia Institute of Technology, Atlanta,
GA 30332, USA}


\begin{abstract}
We describe the effects of disorder on the critical temperature of
$s$-wave superfluids from the BCS to the BEC regime, with direct
application to ultracold fermions. We use the functional integral
method and the replica technique to study Gaussian correlated
disorder due to impurities, and we discuss how this system can
be generated experimentally. In the absence of disorder,
the BCS regime is characterized by pair breaking and
phase coherence temperature scales which are essentially
the same allowing strong correlations between
the amplitude and phase of the order parameter for superfluidity.
As non-pair breaking disorder is introduced the
largely overlapping Cooper pairs conspire to maintain phase
coherence such that the critical temperature remains essentially
unchanged, and Anderson's theorem is satisfied.
However in the BEC regime the pair breaking and phase
coherence temperature scales are very different such that non-pair
breaking disorder can affect dramatically phase coherence, and thus
the critical temperature, without the requirement of breaking
tightly-bound fermion pairs simultaneously.
In this case, Anderson's theorem does not apply, and the
critical temperature can be more easily reduced
in comparison to the BCS limit. Lastly, we find that
the superfluid is more robust against disorder in the intermediate
region near unitarity between the two regimes.
\end{abstract}



\maketitle


\section{Introduction}

Ultracold atoms are special systems for studying superfluid phases
of fermions or bosons at very low temperatures, because of their
unprecedented tunability. In particular, ultracold fermions with
tunable interactions were used to study experimentally the so-called
BCS-to-BEC evolution, where fermion superfluids were investigated as
a function of the interaction parameter (scattering length). The
study of superfluidity in ultracold fermions has additional
promising research directions which include the BCS-to-BEC evolution
in optical lattices~\cite{esslinger-06,ketterle-06,iskin-05}, and
the effects of disorder during the BCS-to-BEC evolution, which would
allow the very important study of the simultaneous effects of
interactions and disorder at zero~\cite{orso-08,shklovskii-08} and
finite temperatures~\cite{sademelo-08,li-10}.

\begin{figure} [htb]
\centerline{\scalebox{0.35}{\includegraphics{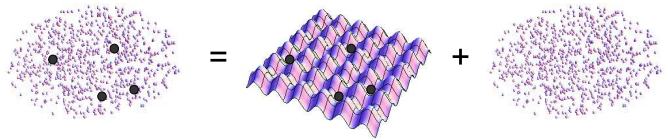} } }
\caption{ \label{fig:one} Schematic illustration of the heavier atom
(e.g. $^{40}$K) impurities trapped in an optical lattice and the
trapped cloud of lighter atoms (e.g. $^{6}$Li) which become
superfluid at low temperatures and low impurity concentration.}
\end{figure}

In ordinary condensed matter (CM) systems the control of
interactions is not possible, and the control of disorder caused by
impurities is very limited because their concentrations can not be
changed at the turn of a knob. However, in ultracold atoms it may be
possible to create disordered impurity potentials, where control
over the impurity concentration and potential could occur in
conjunction with the control over atom-atom interactions. Impurity
type potentials have been very recently created experimentally in
the context of thermometry for spin-dependent optical
lattices~\cite{demarco-10}. However, we envision the creation of
another type of impurity potential where there are two types of
atoms, e.g., a mixture of atoms with unequal masses~\cite{iskin-08}.
The first type of atom (the heavier one) is tightly trapped by an
optical lattice in a low filling fraction configuration, such that
the locations of the atoms for any given realization is unknown
(random). The second type of atom (the lighter one) is trapped by a
harmonic potential, but does not feel the optical lattice potential
of the first type which exists in the same spatial region. In this
case, the lighter atoms only interact with randomly distributed
heavier atoms which are then seen as randomly distributed
impurities, and both the density of impurities (heavier atoms) and
the scattering potential between the impurities (heavier atoms) and
the lighter atoms can be controlled. This kind of situation is
illustrated schematically in Fig.~\ref{fig:one}.

Another way of introducing disorder in ultracold atoms relies on the
use of laser speckles or lasers with incommensurate wavelengths,
which were used to study the phenomenon of Anderson localization in
ultracold Bose atoms~\cite{aspect-08, modugno-08}. The disorder
introduced in this way is very different from impurity disorder in
two important aspects. First, laser speckles or lasers with
incommensurate wavelengths produce a disorder potential landscape
that have many ups and downs ({\it valleys and mountains}) over a
characteristic length scale comparable to the laser wavelength,
while in the case of impurity disorder considered here, the disorder
potential is purely repulsive, that is, has only ups ({\it
mountains}) located at random positions. Second, in laser speckles,
the disorder is exponentially correlated, while in the impurity
problem the disorder is Gaussian correlated. These two major
differences may lead to results that are both quantitatively and
qualitatively different, enforcing the idea that not all types of
disorder produce the same results. Two interesting review articles
have emerged recently covering mostly the effects of disorder in
ultracold Bose atoms~\cite{palencia-10,modugno-10}.

In this manuscript, we describe the critical temperature of three
dimensional (3D) $s$-wave Fermi superfluids from the BCS to the BEC
limit as a function of disorder, which is independent of the
hyperfine states of the atoms and is created by randomly distributed
impurities. However, the effects of disorder during the BCS-to-BEC
evolution in higher angular pairing~\cite{iskin-07} or for
spin-dependent (hyperfine-state-dependent) impurity
potentials~\cite{sademelo-08}  can also be explored.

Our main results are as follows. First, in the BCS limit the
amplitude and the phase of the order parameter are strongly coupled,
such that pair breaking and loss of phase coherence occur
simultaneously. In this case, the critical temperature is
essentially unaffected by weak disorder, since the disorder
potential is not pair-breaking and phase coherence is not easily
destroyed in accordance with Anderson's theorem~\cite{anderson-59}.
Second, in the BEC limit the breaking of local pairs and the loss of
phase coherence occur at very different temperature scales. In this
case, the critical temperature is strongly affected by weak disorder
(in comparison to the BCS regime), since phase coherence is more
easily destroyed without the need to break local pairs
simultaneously, and Anderson's theorem does not apply. Third, we
find that superfluidity is more robust to disorder in the
intermediate region near unitarity between the BCS and BEC regimes.

The rest of the manuscript is organized as follows. In section
\ref{sec:hamiltonian}, we discuss the Hamiltonian for the impurity
problem and the corresponding impurity potential. In section
\ref{sec:effective-action}, we describe the effective action in the
presence of disorder near the critical temperature, where the order
parameter for superfluidity is small. In section
\ref{sec:critical-temperature}, we apply the replica trick to obtain
the critical temperature as a function of disorder and interaction
parameters. We calculate the thermodynamic potential first and then
the number equation in the presence of disorder. In section
\ref{sec:analytical-result}, we present analytical results for weak
disorder both in the BCS and BEC limits. In section
\ref{sec:mean-free-path}, we analyze the mean free paths in limiting
cases and relate them to the parameter describing disorder. Before
concluding, we compare qualitatively speckle and impurity potentials
in section \ref{sec:difference}. Finally, we summarize our
conclusions in section \ref{sec:conclusion}.

\section{Hamiltonian}
\label{sec:hamiltonian}
%
%
In order to study the effects of disorder in Fermi superfluids from
the BCS to the BEC regime we start with the real space Hamiltonian
density for three dimensional $s$-wave superfluids (set $\hbar = 1$)
\begin{equation}
\label{eqn:hamiltonian-real-space}
{\cal H} ({\bf x}) =
\sum_\sigma
\psi^{\dagger}_{\sigma} ({\bf x})
\left(
-\frac{\nabla^2}{2m} - \mu
+ V_{\rm dis} ({\bf x})
\right)
\psi_\sigma ({\bf x}) + {\hat U} ({\bf x}),
\end{equation}
where $\psi^{\dagger}_{\sigma} ({\bf x})$ represents the creation of
fermions with mass $m$ and hyperfine state (spin) $\sigma$ at ${\bf
x}$, $V_{\rm dis} ({\bf x})$ is the disorder potential, and $\mu$ is
the chemical potential. In addition, the interaction term
\begin{equation}
{\hat U} ({\bf x}) = + \int {\mathrm d} x^{\prime}  V ( {\bf x},
{\bf x}^{\prime} ) \psi^{\dagger}_\uparrow ({\bf x}^{\prime})
\psi^{\dagger}_\downarrow ({\bf x}) \psi_{\downarrow}({\bf x})
\psi_{\uparrow} ({\bf x^{\prime}})
\end{equation}
contains the contact interaction potential
$V ({\bf x},{\bf x}^{\prime}) = -g \delta ( {\bf x} - {\bf x}^{\prime} )$.

We choose the disorder potential $V_{\rm dis} ({\bf x})$ to describe
the presence of random impurities and to be independent of the
hyperfine state, a choice that can be easily relaxed. As discussed
in the introduction this choice is quite different from the cases of
disorder introduced by laser speckles or incommensurate lattice
potentials. In the present case, we consider impurity atoms to be
heavier than the other type of atoms in the superfluid state, and it
is also more useful to have the impurity atoms being fermions to
prevent any possibility of low density superfluid behavior, as would
happen for bosons in optical lattices at low filling factors. So for
instance, one could choose $^{40}$K to be the impurity atoms
randomly distributed in a three dimensional lattice that overlaps
with a cloud of $^6$Li atoms, such that $^6$Li and $^{40}$K only
interact with each other without feeling the presence of the other
atom's trapping potential. This means that the lighter atoms (e.g.
$^6$Li) do not feel the presence of the optical lattice trapping the
heavier atoms (e.g. $^{40}$K), and conversely the heavier atoms do
not feel the presence of the harmonic potential confining the
lighter atoms, but the lighter and heavier atoms interact with each
other. In this case, the interaction between the heavier impurity
atoms and the lighter atoms can be expressed by an effective
impurity potential
\begin{equation}
V_{\rm dis} ({\bf x})
=
\sum_i v_d F({\bf x} - {\bf r}_i),
\end{equation}
where ${\bf r}_i$ are the locations of the impurities, and $F ({\bf
R}) = \pi^{-3/2} \ell_d^{-3} \exp(- R^2/\ell_d^2)$ is a Gaussian of
width $\ell_d$ describing the range of the impurity potential.
Notice that $v_d$ is a measure of the interaction between heavier
impurity atoms and lighter atoms, and $F ({\bf x} - {\bf r}_i)$ is a
measure of the square of the Wannier wavefunction $\vert \varphi
({\bf x} - {\bf r}_i) \vert^2$ of the heavier impurity atom. If the
interactions between every heavier impurity atom and every lighter
atom are the same and always repulsive, the effective impurity
potential is smooth and strictly repulsive.

The  average over impurity positions leads to the correlator
\begin{equation}
\langle V_{\rm dis}
({\bf x}) V_{\rm dis} ({\bf x}^\prime) \rangle = \kappa K ({\bf x} -
{\bf x}^\prime),
\end{equation}
where $\kappa = n_i v_d^2$ with $n_i$ being the impurity
concentration, and the Kernel
\begin{equation}
K ({\bf x} - {\bf x}^\prime)
=
\frac{1}
{(2\pi)^{3/2} \ell_d^3}
\exp \left[ - ({\bf x} - {\bf x}^\prime)^2/2 \ell_d^2 \right].
\end{equation}
By inspection, only in the limit of $\ell_d \to 0$,
the correlator becomes $\langle V_{\rm dis}
({\bf x}) V_{\rm dis} ({\bf x}^\prime) \rangle = \kappa \delta ({\bf
x} - {\bf x}^\prime)$, reflecting white-noise correlations.

Given that the Hamiltonian is fully described, we discuss next the
effective action of the system, and the corresponding approximations
to obtain the critical temperature of superfluid fermions from the
BCS to the BEC regime in the presence of disorder.

\section{Effective action in the presence of disorder}
\label{sec:effective-action}

The effective action during the evolution from BCS to BEC
superfluidity and the calculation of the critical temperature in the
absence of disorder were obtained long ago~\cite{sademelo-93}, and
are briefly reviewed here to provide background for the derivation
of the effective action in the presence of disorder.

We begin our discussion by recalling that the effective action in
the absence of disorder can be derived by introducing the pairing field
$
\Delta (x)
=
\langle \psi_{\downarrow} (x) \psi_{\uparrow} (x) \rangle
$
as a thermal average, where  $x = ({\bf x}, \tau)$, with $\tau$
being the imaginary time.  Upon integration of the residual
fermionic degrees of freedom, the resulting action up to Gaussian order
in the pairing field~\cite{sademelo-93} has two parts
\begin{equation}
S_{\rm eff} = S_0 + S_{\rm Gau}.
\end{equation}
The first term describes the action of unbound
fermions
\begin{equation}
S_0 = -2 \sum_{\bf k} \ln \left[ 1 + {\mathrm e}^{- \xi({\bf
k})/T}\right],
\end{equation}
with $\xi ({\bf k}) =
\epsilon ({\bf k}) - \mu$, where $\epsilon ({\bf k}) = k^2/2m$ is
the kinetic energy of a fermion of momentum ${\bf k}$.
The second term is the Gaussian action
\begin{equation}
S_{\rm Gau} = T^{-1} \sum_{q} {\bar \Delta} (q) \Gamma^{-1} (q)
\Delta (q)
\end{equation}
due to the pairing field $\Delta (q)$ and its complex conjugate
${\bar \Delta} (q)$ in momentum space, the former of which is just
the Fourier transform of $\Delta (x)$. Here,
\begin{equation}
\Gamma^{-1} (q) = \frac{V}{g} + \sum_{\bf k} \frac{ X_1 + X_2}{2(
{\mathrm i} q_{\ell} - \xi_1 - \xi_2)}
\end{equation}
is the fluctuation propagator, and $q = ({\bf q},{\mathrm i}
q_{\ell})$ is the four-momentum with $q_{\ell} =  2\pi \ell T$ being
the bosonic Matsubara frequencies. Here, the function $ X_1 = \tanh
\left[ \xi_1 /2T \right] $ describes the occupation of fermions with
energy $ \xi_1 = \xi ( {\bf k} - {\bf q}/2 ) $, and the function $
X_2 = \tanh \left[\xi_2/2 T \right] $ describes the occupation of
fermions with energy $ \xi_2 = \xi ( {\bf k} + {\bf q}/2 ). $

To derive the effective action for a fixed configuration of
disorder, we define the local chemical potential $\mu ({\bf x}) =
\mu - V_{\rm dis} ({\bf x})$. This corresponds to the local density
approximation and treats the effects of the disorder potential
semiclassically. In the case of the impurity potential originated by
heavier atoms randomly distributed as described above, this is
expected to be a reasonable approximation, as the impurity potential
is purely repulsive and non-confining such that bound states of the
impurity potential do not exist. The situation for speckle
potentials or other random potentials that possess bound states is
quite different, and will be briefly discussed later.

Within the local density approximation, the effective action in the
presence of impurity disorder becomes
\begin{equation}
S_{\rm eff}[V_{\rm dis}] = S_0 [V_{\rm dis}] + S_{\rm pair} [V_{\rm
dis}],
\end{equation}
where $S_0 [V_{\rm dis}]$ is the action of unbound fermions
in the presence of disorder given by
\begin{equation}
\label{eqn:effective-action-zero} S_0 [V_{\rm dis}] =  - 2 \int
\frac{{\mathrm d}{\bf x}}{V} \sum_{\bf k} \ln \left[ 1 + {\mathrm e
}^{-\xi ({\bf k}, {\bf x})/T} \right],
\end{equation}
with $\xi ({\bf k}, {\bf x}) = \epsilon ({\bf k}) - \mu({\bf x})$.
The second contribution to $S_{\rm eff}$ has two terms, that is
$S_{\rm pair} = S_{\rm Gau} + S_4$. The first term corresponds to
Gaussian pairing fluctuations
\begin{equation}
\label{eqn:effective-action-gaussian} S_{\rm Gau} [ V_{\rm dis} ] =
\int  \frac{{\mathrm d}{\bf x}}{TV} \sum_{q} {\bar \Delta} (q, {\bf
x} ) \Gamma^{-1} (q, V_{\rm dis}) \Delta (q, {\bf x} ),
\end{equation}
where $\Delta (q, {\bf x})$ is the pairing field, 
and
\begin{equation}
\label{eqn:pair-correlation} \Gamma^{-1} (q, V_{\rm dis}) =
\frac{V}{g} + \sum_{\bf k} \frac{ {\widetilde X}_1 + {\widetilde
X}_2 } {\left[ 2 ({\mathrm i} q_{\ell} - \xi_1 - \xi_2  - 2 V_{\rm
dis} ({\bf x})) \right] }
\end{equation}
is the pair correlation function in the presence of disorder. Here,
the atom-atom interaction $g$ can be expressed in terms of the
scattering length $a_s$ as
\begin{equation}
\frac{V}{g}
=
- \frac{mV}{4\pi a_s}
+ \sum_{\bf k} \frac{1}{2 \epsilon_{\bf k}}.
\end{equation}
In addition, the notation $q = ( {\bf q}, {\mathrm i} q_{\ell} )$
represents the four-momentum, the function ${\widetilde X}_1 = \tanh
\left[ \left( \xi_1  + V_{\rm dis} ({\bf x}) \right)/2T \right] $
describes the occupation of fermions with energy $\xi_1 = \xi ( {\bf
k} - {\bf q}/2 )$, and the function ${\widetilde X}_2 =  \tanh
\left[ \left( \xi_2  + V_{\rm dis} ({\bf x}) \right) /2 T \right] $
describes the occupation of fermions with energy $\xi_2 = \xi ( {\bf
k} + {\bf q}/2 )$.

The interaction between the pairing fields is described by the action
\begin{equation}
S_4 = (TV)^{-1} \int {\mathrm d} {\bf x} {\mathrm d} \tau
\frac{b}{2} \vert \Delta ({\bf x}, \tau) \vert^4,
\end{equation}
where we used the definition
$
\Delta ({\bf x}, \tau)
=
\Delta ({\bf q} = 0, {\bf x}, \tau).
$
Here, the coefficient
\begin{equation}
b = \sum_{\bf k} \left[ \frac{X}{4\xi_ {\bf k}^3} - \frac{Y}{8T
\xi_{\bf k}^2} \right]
\end{equation}
represents the strength of the effective interaction between pairing
fields, and the notations $X = \tanh(\xi_{\bf k}/2T) $ and $Y = {\rm
sech}^2 (\xi_{\bf k}/2T)$ have been used.

An expansion of the pair correlation function $\Gamma^{-1} (q,
V_{\rm dis})$ in the limit of low energy (small ${\mathrm i}
q_{\ell}$, with the analytic continuation ${\mathrm i} q_{\ell} \to
\omega + {\mathrm i} \delta$), long-wavelength (small $\vert {\bf q}
\vert$), and small disorder potential leads to
\begin{equation}
\label{eqn:pair-correlation-expansion} \Gamma^{-1} (q, V_{\rm dis})
= a + c \frac{\vert {\bf q} \vert^2}{2m} + e V_{\rm dis} ({\bf x}) +
d \omega.
\end{equation}
The coefficients of the Taylor expansion are
\begin{equation}
\label{eqn:a}
a (\mu, T)
=
-\frac{m V}{4\pi a_s}
+ \sum_{\bf k} \left[ \frac{1}{ 2 \epsilon_{\bf k} } -
\frac{X} { 2\xi_{\bf k} } \right],
\end{equation}
describing the order parameter equation in the absence of disorder
when $a(\mu, T) = 0$,
\begin{equation}
c (\mu, T) = \sum_{\bf k} \left[ \frac{X}{8\xi_{\bf k}^2} -
\frac{Y}{16\xi_{\bf k} T} + \frac{XY}{T^2} \frac{k_z^2} {16 m
\xi_{\bf k} } \right],
\end{equation}
describing the kinetic energy of the pairing field, $d (\mu, T) =
d_R + {\mathrm i} d_I$, with $d_R = \sum_{\bf k} {X}/{4 \xi_{\bf
k}^2} $ and $d_I = \left[ \pi N(\epsilon_F) \sqrt{\mu}/(8 T
\sqrt{\epsilon_F} ) \right] \Theta (\mu)$ describing the frequency
dependence, and $e (\mu, T)  = -\partial a/\partial \mu \vert_T$
describing the strength of effective disorder potential.

It is very convenient to rewrite the effective action for the
pairing field $S_{\rm pair}$ into a more familiar form. Thus, next,
we put Eq.~(\ref{eqn:pair-correlation-expansion}) back into the
action of Eq.~(\ref{eqn:effective-action-gaussian}), set $\vert {\bf
q} \vert^2 \to - \nabla^2$, $\Delta (q, {\bf x}) \to \Delta ({\bf q}
= 0, {\rm i}q_{\ell}, {\bf x})$, and scale the pairing field
$\Delta/\sqrt{d_R} \to \Psi$ such that the coefficient of the real
part of $d \omega$ becomes one. A subsequent Fourier transformation
into imaginary time, and the inclusion of the fourth order term in
$\Psi$ leads to the pairing field action
\begin{equation}
S_{\rm pair} [V_{\rm dis}] = S_{\rm Gau} [V_{\rm dis}] + S_4 [V_{\rm
dis}],
\end{equation}
where $S_{\rm Gau} [V_{\rm dis}]$ is defined in
Eq.~(\ref{eqn:effective-action-gaussian}) and
\begin{equation}
S_4 [V_{\rm dis}] = (TV)^{-1} \int {\mathrm d} {\bf x} {\mathrm d}
\tau \frac{g^*}{2} \vert \Psi ({\bf x}, \tau)\vert^4,
\end{equation}
where $g_{*} = b/d_R^2$ is the effective interaction.
In order to explicit out dissipative and non-dissipative
temporal fluctuations, the pairing action is rewritten as
\begin{equation}
\label{eqn:effective-action-pair} S_{\rm pair} [{\bar \Psi}, {\Psi},
V_{\rm dis}] = (TV)^{-1} \int {\mathrm d} {\bf x} {\mathrm d} \tau
{\cal L}_{\rm pair},
\end{equation}
where ${\cal L}_{\rm pair} = {\cal L}_{ND} + {\cal L}_{D}$ is the
Lagrangian density that contains two contributions. The first term
is a non-dissipative part
\begin{equation}
\label{eqn:lagrangian-non-dissipative}
{\cal L}_{ND} (x, \tau) =
{\bar \Psi} \left[ \partial_{\tau} -
\frac{\nabla^2}{2 m_{*}} - \mu_{*} + \gamma V_{\rm dis} ({\bf x})
\right] \Psi + \frac{g_{*}}{2} \vert \Psi \vert^4,
\end{equation}
and describes a generalized Gross-Pitaeviskii Lagrangian for the
scaled pairing field $\Psi = \Psi ({\bf x}, \tau)$. The term
containing the effective mass $m_{*} = d_R m/c$ is the kinetic
energy of the pairing field, $\mu_{*} = - a/d_R$ plays the role of
the pairing field chemical potential, and $\gamma V_{\rm dis} ({\bf
x})$ is the effective disorder potential, with $\gamma = e/d_R$. The
second term is the dissipative contribution that comes from the term
containing ${\mathrm i} d_I \omega$ and reflects the decay of
fermion pairs into unbound fermions (Landau damping), which in
imaginary time takes the Caldeira-Leggett form
\begin{equation}
\label{eqn:lagrangian-dissipative} {\cal L}_{D} (x, \tau) =
\frac{\lambda}{2\pi}\int {\mathrm d} \tau^{\prime} \frac {\vert \Psi
({\bf x}, \tau) - \Psi ( {\bf x}, \tau^{\prime}) \vert^2} {(\tau -
\tau^{\prime})^2},
\end{equation}
where $\lambda = d_I/d_R$. In our present discussion $\lambda \ne 0$
for $\mu < 0$ and $\lambda = 0$ for $\mu > 0$, since $d_I$ is
non-zero for positive $\mu$, but vanishes for negative $\mu$. This
indicates that on the BCS side $(\mu > 0)$ the lifetime of pairs is
short (pairs break) near the critical temperature, while on the BEC
side $\mu < 0$ the lifetime of the pairs is infinite, or better
said, very large, such that stable fermion pairs exist even above
the critical temperature.

It is important to emphasize that the long-wavelength, low-frequency
effective action derived in Eq.~(\ref{eqn:effective-action-pair})
is valid for the entire evolution from BCS to BEC superfluidity near
the critical temperature $T_c$, so long as the local density approximation
is applicable. Now that we have derived the effective action for
the pairing field in the presence of disorder, i.e.,
a purely repulsive random impurity potential, we discuss
next the self-consistency equations that output the critical
temperature as a function of disorder for a given interaction
parameter.

\section{The critical temperature in the presence of disorder}
\label{sec:critical-temperature}

In order to determine the critical temperature in the presence of
disorder, we need to derive the corresponding order parameter and
number equations. For this purpose, a natural choice of
dimensionless parameters are $1/(k_F a_s)$ for interactions, $\eta =
\kappa n_F/\epsilon_F^2$ for disorder, and ${\widetilde T} =
T/\epsilon_F$ for temperature, where $k_F$ is the Fermi momentum,
$n_F = k_F^3/3\pi^2$ is the fermion density and $\epsilon_F =
k_F^2/2m$ is the Fermi energy. Notice that our dimensionless
parameter $\eta$ that characterizes the degree of disorder is
directly obtained from the original Hamiltonian density in
Eq.~(\ref{eqn:hamiltonian-real-space}) and from the disorder
correlator $\langle V_{\rm dis} ({\bf x}) V_{\rm dis} ({\bf
x}^\prime) \rangle$, and is thus an input parameter of the theory.
Since $\kappa = n_i v_d^2$, and $v_d$ has dimensions of energy times
volume, it can be written as the product of the impurity volume
$\ell_d^3$ and the characteristic amplitude ${\widetilde v}_d$ of
the disorder potential $V_{\rm dis} ({\bf x})$, i.e. $v_d =
{\widetilde v}_d \ell_d^3$. Therefore, we may write $\eta = (n_i
\ell_d^3)(n_F \ell_d^3)({\widetilde v}_d/\epsilon_F)^2$, which
reveals that $\eta$ is small when the range of the impurity
potential $\ell_d$ is much smaller than the average separation
between impurities $n_i^{-1/3}$ ($n_i^{-1/3} \ell_d \ll 1$) or
between fermions $n_F^{-1/3}$ ($ n_F^{-1/3} \ell_d \ll 1$), and also
when the amplitude of the disorder potential ${\widetilde v}_d$ is
much smaller than the Fermi energy $\epsilon_F$ (${\widetilde
v}_d/\epsilon_F \ll 1$). In some of the older literature discussing
the effects of disorder for non-interacting fermions, the
dimensionless parameter is typically chosen to be $1/(k_F \ell)$,
where $\ell$ is the mean free path. Note that is a derived quantity,
and not an input parameter of the theory. Even though the mean free
path is a useful concept, when the Fermi system is strongly
interacting it becomes increasingly more difficult to define it, and
also to calculate it. Thus we prefer to use our input parameter
$\eta$ to describe the degree of disorder of the system. Later in
our discussion, we will return to the connection between our
dimensionless parameter $\eta$, and the mean free path of paired and
unpaired fermions.
%
%

The order parameter equation can be immediately read from
Eq.~(\ref{eqn:lagrangian-non-dissipative}) and is given by the
condition
\begin{equation}
\label{eqn:order-parameter-disorder} \mu_{*} ( {\widetilde T},
1/(k_F a_s) ) = 0
\end{equation}
corresponding precisely to the order parameter equation in the absence
of disorder $a (\mu, T) = 0$ given in Eq.~(\ref{eqn:a}),
indicating that this equation is not explicitly affected by
weak disorder, as required by Anderson's theorem~\cite{anderson-59}.
However, the determination of the
critical temperature $T_c (\eta)$ and the chemical potential
$\mu (\eta)$ as a function of dimensionless disorder $\eta$ for fixed
scattering parameter $1/(k_F a_s)$ requires the simultaneous
solutions of the order parameter
Eq.~(\ref{eqn:order-parameter-disorder}) and the number equation,
which is discussed next.

To determine the number equation in the presence of disorder, it is
necessary to average the thermodynamic potential $\Omega (V_{\rm
dis})$ rather than the partition function~\cite{belitz-94}. Thus, to
compute the disorder average $\langle ... \rangle = \int {\cal D} V
P[V] (...)$, where the probability measure is
\begin{equation}
 P [V]
= \exp \left[ - \frac{1}{\kappa} \int {\mathrm d} {\bf x} {\mathrm
d} {\bf x}^\prime V_{\rm dis} ({\bf x}) K^{-1} ( {\bf x} - {\bf
x}^\prime ) V_{\rm dis} ({\bf x}^\prime) \right],
\end{equation}
we just need the disorder correlator $K ({\bf x} - {\bf x}^\prime)$.
The thermodynamic potential for a fixed configuration is $\Omega
(V_{\rm dis})  = - T \ln Z (V_{\rm dis}) $, and $Z (V_{\rm dis})$
can be expressed as the product of two contributions
\begin{equation}
Z (V_{\rm dis}) = Z_0 (V_{\rm dis}) \times Z_{\rm pair} (V_{\rm
dis}).
\end{equation}
The first term is the partition function for unbound fermions
\begin{equation}
Z_0 (V_{\rm dis})
=
\exp \left[ - S_0 (V_{\rm dis}) \right]
\end{equation}
and the second term is the partition function for paired fermions
\begin{equation}
Z_{\rm pair} (V_{\rm dis}) = \int {\cal D} [\bar \Psi, \Psi] \exp
\left[- S_{\rm pair} ( {\bar \Psi}, {\Psi}, V_{\rm dis}) \right].
\end{equation}
Thus, $\Omega (V_{\rm dis}) = \Omega_0 (V_{\rm dis})  +
\Omega_{\rm pair} (V_{\rm dis})$, where the unbound fermion
thermodynamic potential is $\Omega_0 (V_{\rm dis}) = - T \ln Z_0
(V_{\rm dis}) = T S_0 (V_{\rm dis})$, while the contribution due to
paired fermions is $\Omega_{\rm pair} (V_{\rm dis}) = - T \ln Z_{\rm
pair} (V_{\rm dis})$. However, the calculation of the average
$\langle \Omega \rangle = - T \langle \ln Z \rangle $ can be
performed by using the standard replica trick~\cite{belitz-94} for
statistical averages, where
\begin{equation}
\langle \ln Z \rangle  = {\lim}_{M \to 0} \ln \langle Z^M
\rangle^{1/M}.
\end{equation}

\begin{figure} [htb]
\centering
\begin{tabular}{cc}
\epsfig{file = 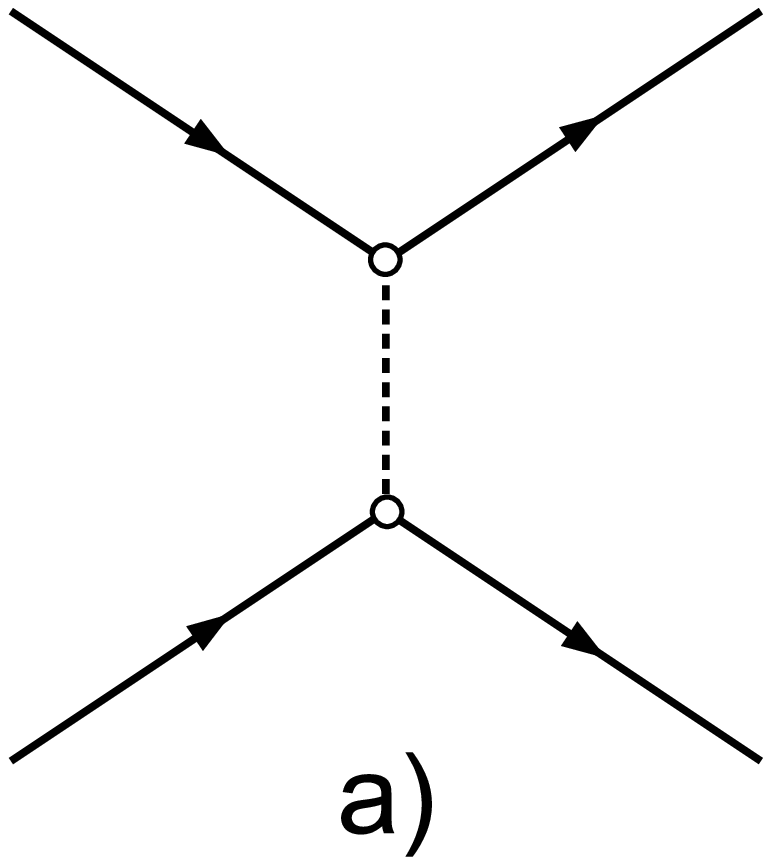, width=0.50\linewidth,clip=} 
\epsfig{file = 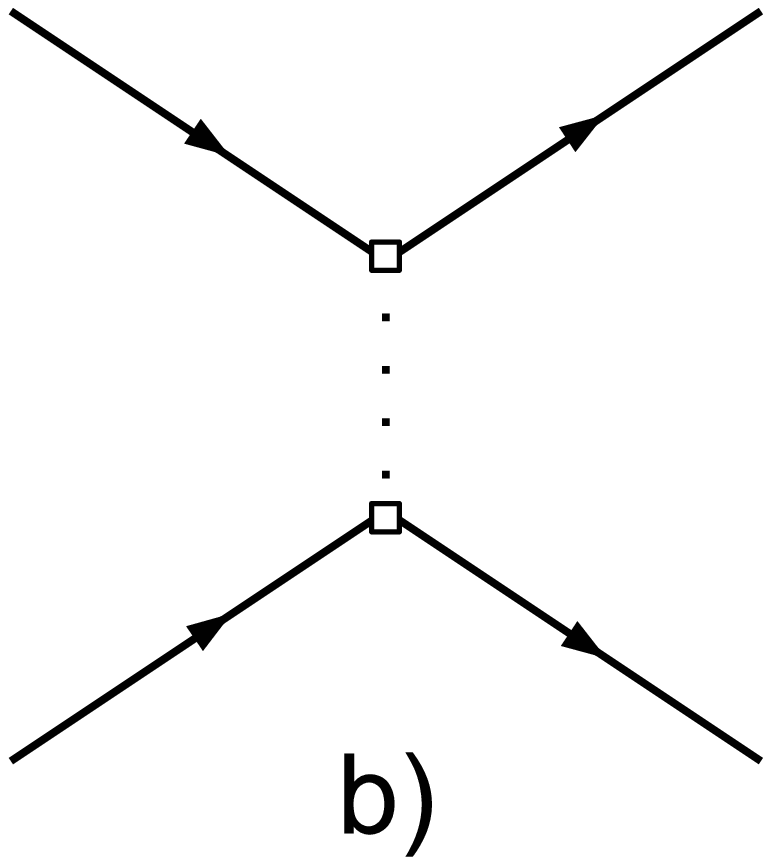, width=0.50\linewidth,clip=}
\end{tabular}
\caption{ \label{fig:two} The two types of interaction lines between
the replicated paired fields. The diagram in a) is the direct
intra-replica interaction, containing $g_*$, while the diagram in b)
is the inter-replica interaction containing $\kappa$. }
\end{figure}

This trick was first introduced in the context of
spin glasses~\cite{anderson-75},
and can be applied to both the unbound and paired fermion
parts of the effective action. Here, however, we discuss explicitly
the application of the replica trick only to the more complex case
of the pairing field, and just quote the result for the
thermodynamic potential of unbound fermions.
In the case of the pairing field, the replicated ($M$-copies)
partition function is
\begin{equation}
Z_{\rm pair}^{M} = \int \prod_{i=1}^{M} {\cal D} \left[{\bar
\Psi}_i, \Psi_i \right] \exp \left\{ - \sum_{i = 1}^M S_{\rm pair}
\left[ {\bar \Psi}_i, \Psi_i, V_{\rm dis} \right] \right\},
\nonumber
\end{equation}
with $S_{\rm pair}$ defined in
Eq.~(\ref{eqn:effective-action-pair}). Taking the configurational
average $\langle Z_{\rm pair}^M \rangle$ with the probability
measure $P \left[ V_{\rm dis} \right]$ amounts solely to a Gaussian
integral over $V_{\rm dis}$, which can be easily performed leading
to
\begin{equation}
\label{eqn:z-pair} \langle Z_{\rm pair}^M \rangle = \int {\cal
D}_{\Psi} \exp \left\{ -\left[ \sum_{i = 1}^M S_{\rm pair} (i,0) +
\sum_{i, j = 1}^M S_{\rm dis} (i, j) \right] \right\}.
\end{equation}
Here, we used the notation ${\cal D}_{\Psi} = \prod_{i=1}^{M} {\cal
D} \left[{\bar \Psi}_i, \Psi_i \right] $ with $ S_{\rm pair} (i, 0)
= S_{\rm pair} \left[ {\bar \Psi}_i, \Psi_i, V_{\rm dis} = 0
\right]$ corresponding to the pair action without disorder
and with the replacement of $\Psi \to \Psi_i$
in Eq.~(\ref{eqn:effective-action-pair}).
The second term in Eq.~(\ref{eqn:z-pair})
corresponds to a density-density interaction
between the replicated fields given by
\begin{equation}
S_{\rm dis} (i,j) = - \frac{\kappa_{\rm pair}}{2} \sum_{ij} \int T
{\mathrm d} \tau {\mathrm d} {\bf x} {\mathrm d} {\bf x}^\prime
\rho_i (x) K ({\bf x} - {\bf x}^\prime) \rho_j (x^\prime) \nonumber
\end{equation}
where $\kappa_{\rm pair} = \gamma^2 \kappa$. The densities $\rho_i
(x) = {\bar \Psi}_i ({\bf x}, \tau){\Psi}_i ({\bf x}, \tau)$ and
$\rho_j (x^\prime) = {\bar \Psi}_j ({\bf x}^\prime, \tau){\Psi}_j
({\bf x}^\prime, \tau)$ appear for equal time $\tau$, while $K({\bf
x} - {\bf x}^\prime)$ is the disorder correlator defined previously.
Thus, the replica trick transforms the problem of the pairing field
$\Psi$ in a disorder potential, into the problem of interacting
replicated pairing fields $\Psi_i$ with interaction strength
$-(\kappa_{\rm pair}/2)  K({\bf x} - {\bf x}^\prime)$, in addition
to the standard direct interaction term characterized by the
interaction $(g^*/2) \, \delta({\bf x} - {\bf x}^\prime)$. There are
now two types of interactions between the replicated pairing fields
as indicated in Fig.~{\ref{fig:two}}.

\begin{figure} [htb]
\centering
\begin{tabular}{cc}
\epsfig{file = 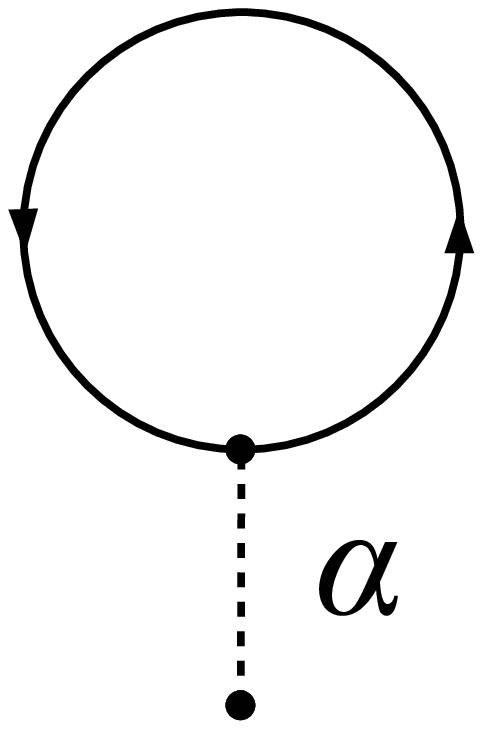, width=0.30\linewidth,clip=} 
\epsfig{file = 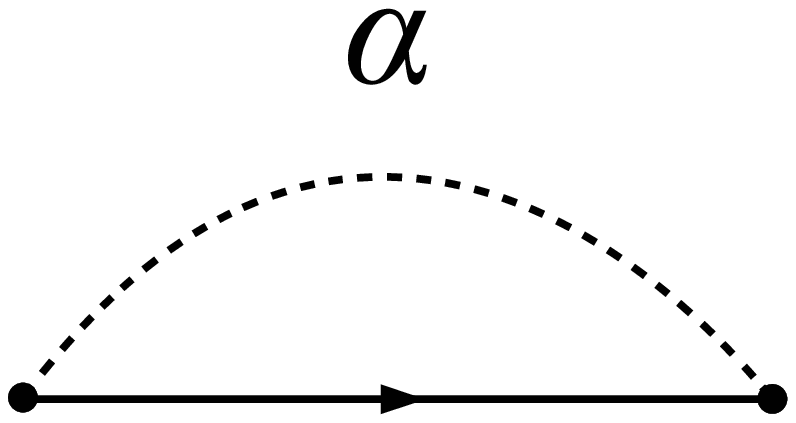, width=0.30\linewidth,clip=}
\end{tabular}
\caption{ \label{fig:three} Two possible first-order self-energy
diagrams side by side for the each interaction line of type $\alpha$. Here,
$\alpha$ describes either the intra-replica or inter-replica
interaction line.}
\end{figure}

\begin{figure} [htb]
\centering
\begin{tabular}{c}
\epsfig{file = 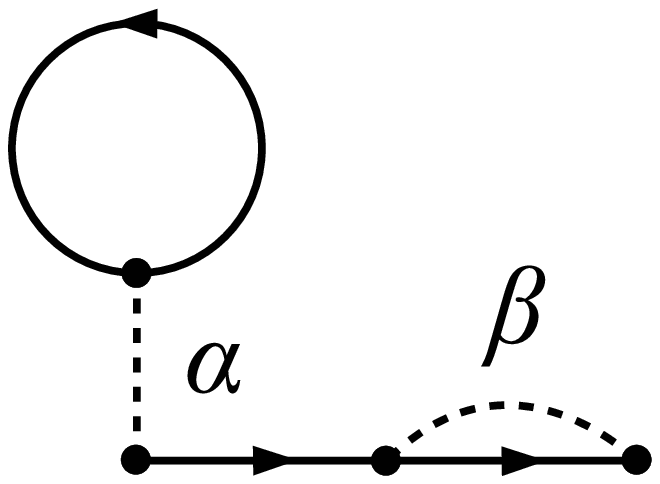, width=0.40\linewidth,clip=} 
\epsfig{file = 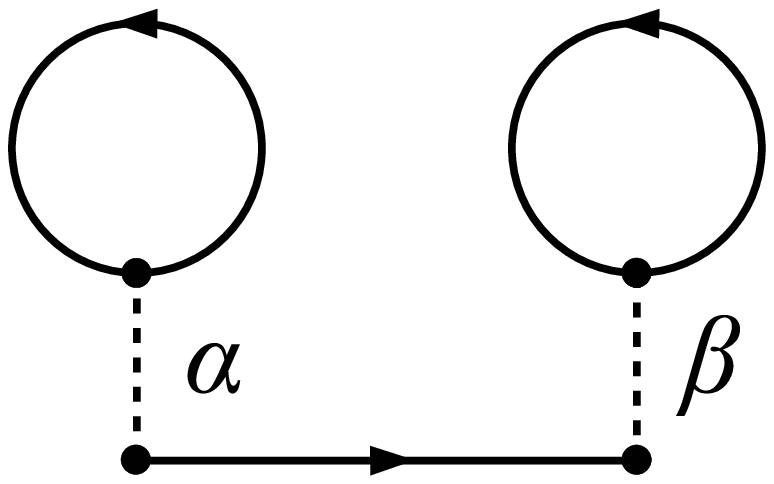, width=0.40\linewidth,clip=} \\
\epsfig{file = 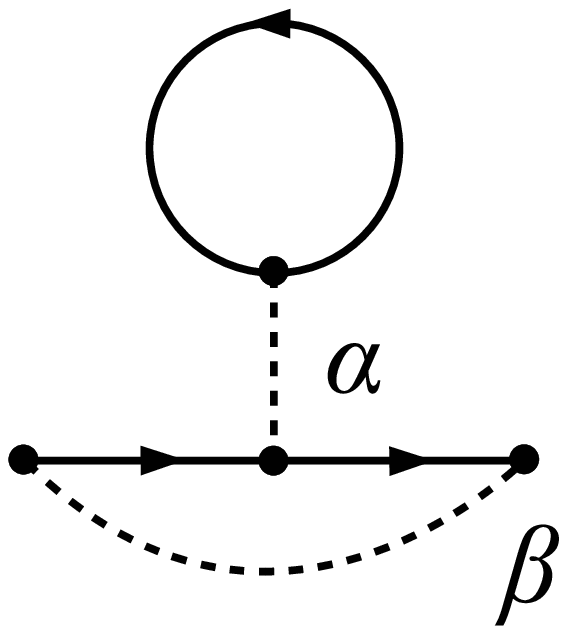, width=0.40\linewidth,clip=} 
\epsfig{file = 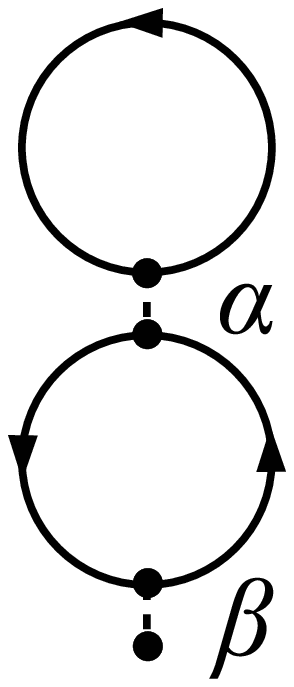, width=0.40\linewidth,clip=} \\
\epsfig{file = 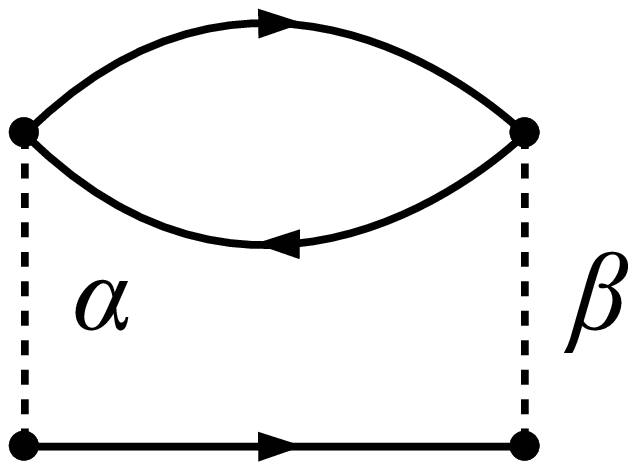, width=0.40\linewidth,clip=} 
\epsfig{file = 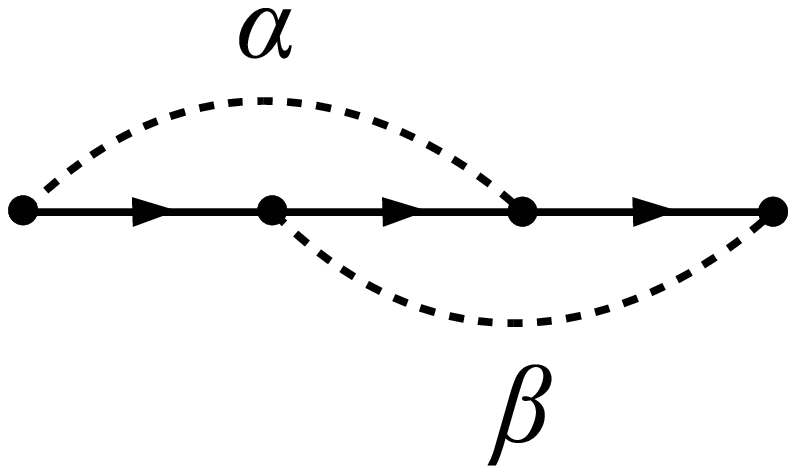, width=0.40\linewidth,clip=} \\
\epsfig{file = 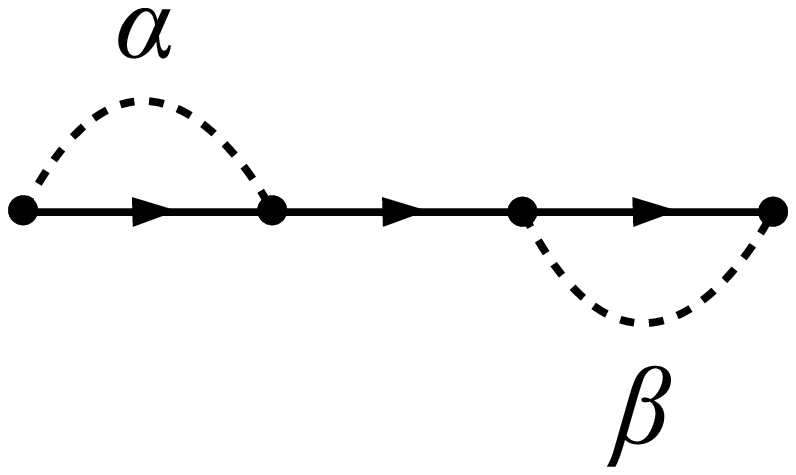, width=0.40\linewidth,clip=} 
\epsfig{file = 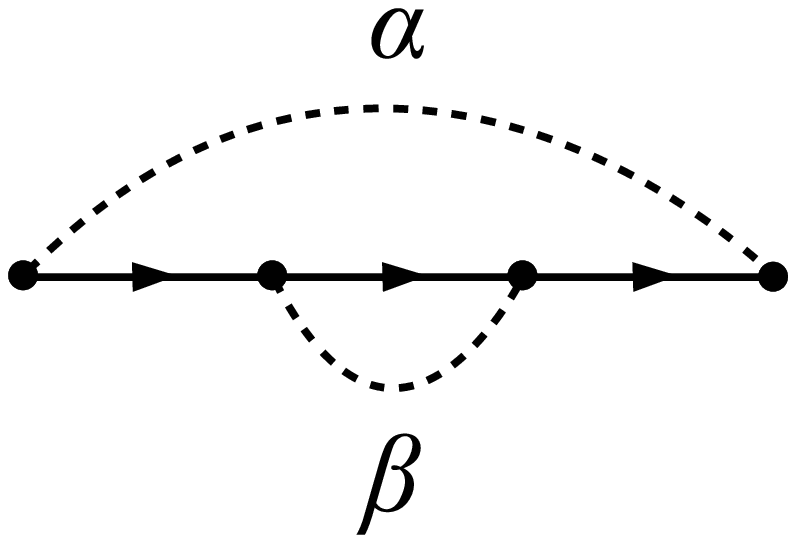, width=0.40\linewidth,clip=}
\end{tabular}
\caption{ \label{fig:four} Eight possible second order self-energy
diagrams side by side for fixed interaction lines labeled $\alpha$
and $\beta$. Here, $\alpha$ and $\beta$ describe either an
intra-replica or inter-replica interaction line.}
\end{figure}

The calculation of the averaged thermodynamic potential for paired
fermions
\begin{equation}
\langle \Omega_{\rm pair} \rangle = - T \lim_{M \to 0} \ln \langle
Z_{\rm pair}^M \rangle^{1/M}
\end{equation}
can now be performed perturbatively as a function of the disorder
parameter $\kappa_{\rm pair}$ and the interaction $g_*$. For this
purpose it is better to work in four-momentum space $q = ({\bf q},
{\mathrm i} q_{\ell})$ and obtain the replica Green's function
\begin{equation}
\label{eqn:replica-green-function}
G_{\rm pair}^{-1} (q) = -{\mathrm
i} q_{\ell} + E_{\rm pair} (q),
\end{equation}
where $ E_{\rm pair} (q) = \lambda \vert q_\ell \vert +\vert{\bf
q}\vert^2/2m_* -\mu_* +\Sigma_{\rm pair} (q). $ Here, $\Sigma_{\rm
pair}  (q)$ is the Dyson's self-energy for paired fermions
containing all possible Feynman diagrams to the desired order in
powers of $g_*$ and $\kappa_{\rm pair}$. These types of diagrams are
standard in diagrammatic perturbation theory~\cite{simon-08}. For
instance, to first order in $g_*$ and $\kappa_{\rm pair}$ the
self-energy has only $2 \times 2 = 4$ Feynman diagrams as shown in
Fig.~\ref{fig:three}, while to second order in $g_*$ and
$\kappa_{\rm pair}$ there are $8 \times 4 = 32$ diagrams as seen in
Fig.~\ref{fig:four}.

%


%


%
\begin{figure} [htb]
\centering
\begin{tabular}{c}
\epsfig{file = 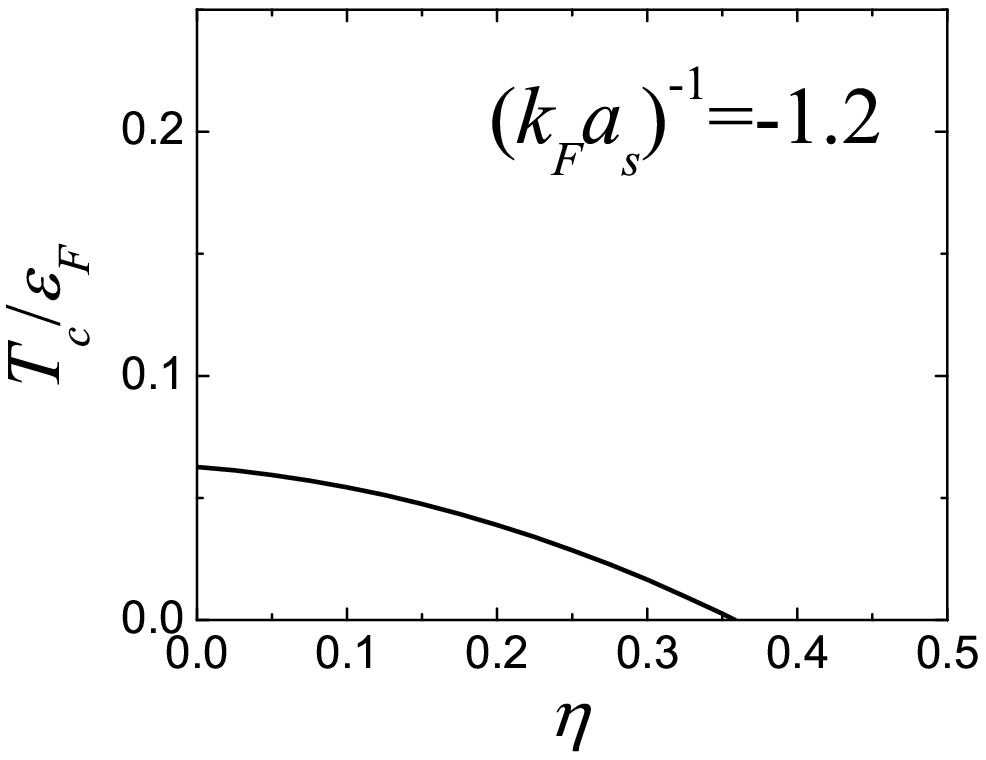, width=0.60\linewidth,clip=} \\
\epsfig{file = 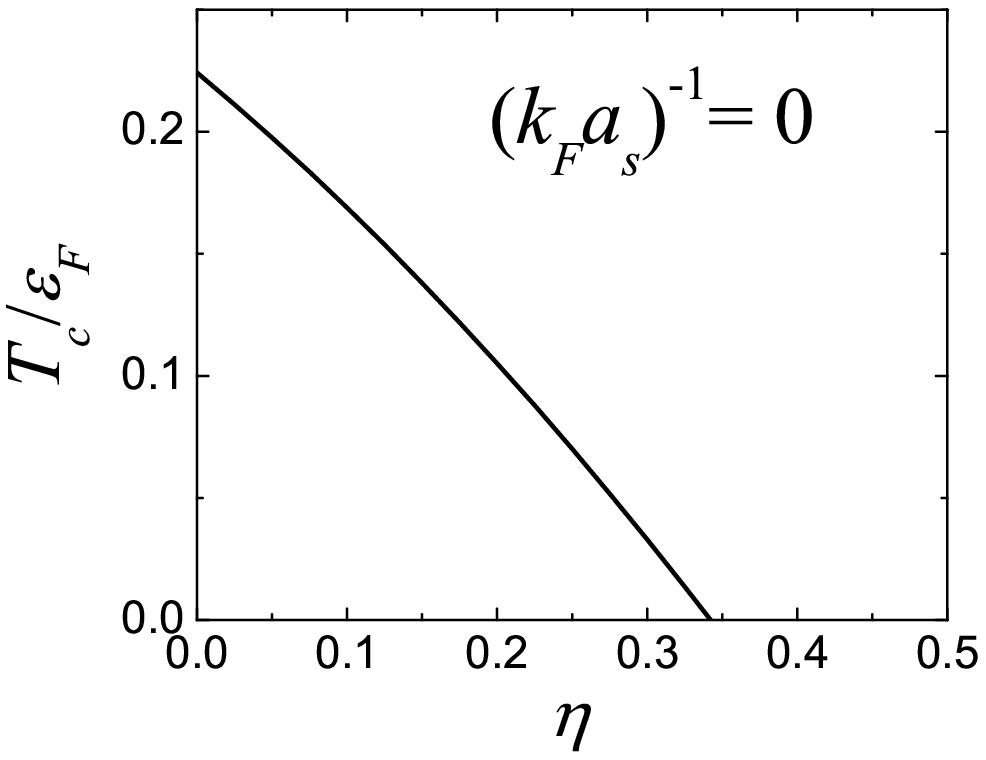, width=0.60\linewidth,clip=} \\
\epsfig{file = 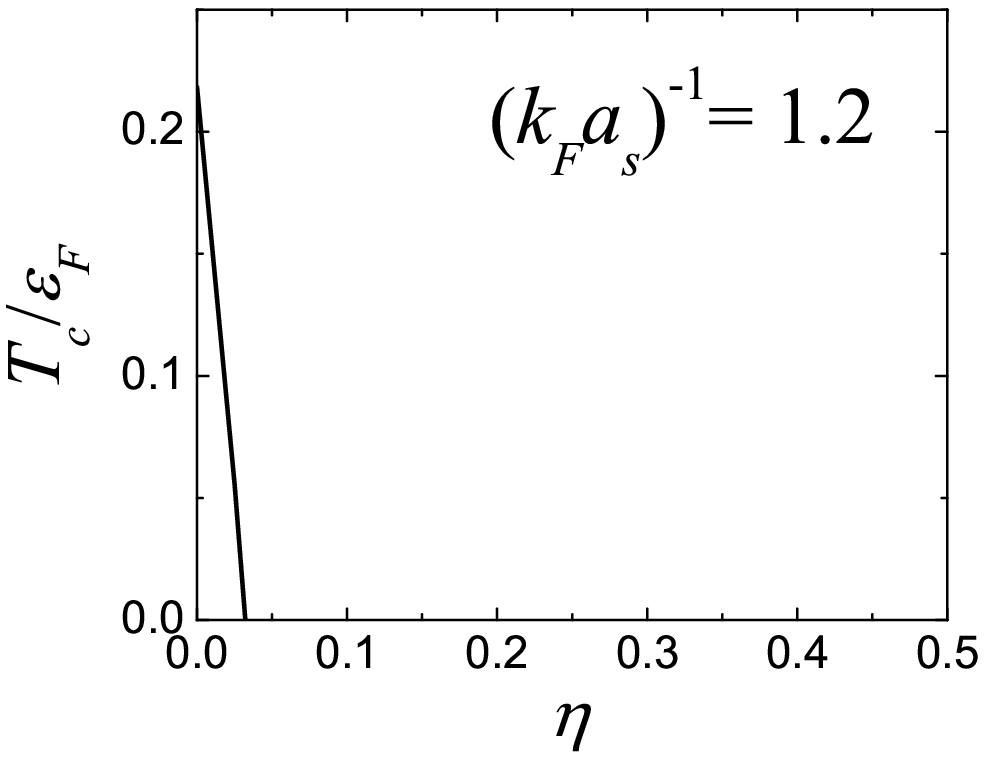, width=0.60\linewidth,clip=}
\end{tabular}
\caption{ \label{fig:five} The critical
temperature $T_c/\epsilon_F$ versus dimensionless disorder
parameter $\eta$ for different values of
$(k_F a_s)^{-1} = -1.2, 0, 1.2$, respectively.}
\end{figure}

The self-energy can be written as the expansion
$\Sigma_{\rm pair} = \Sigma_{\rm pair}^{(1)} + \Sigma_{\rm
pair}^{(2)} + ...$, where
$\Sigma_{\rm pair}^{(1)}$ corresponds to the diagrams
in Fig.~\ref{fig:three}, and  $\Sigma_{\rm pair}^{(2)}$ to the diagrams
in Fig.~\ref{fig:four}.
%
%
%
%
The replicated partition function is $ \langle Z_{\rm pair}^M
\rangle = \left[ \prod_{q} G_{\rm pair}^{-1}(q)/T \right]^{-M} $
leading to the pair thermodynamic potential $ \langle \Omega_{\rm
pair} \rangle = T \sum_{q} \ln \left[ G_{\rm pair}^{-1} (q) / T
\right]$ from which the contribution to the number of paired
fermions can be obtained via the thermodynamic identity $ N_{\rm
pair} = - \partial \langle \Omega_{\rm pair} \rangle /
\partial{\mu}.
$

The disorder averaged thermodynamic potential is
$
\langle \Omega
\rangle = \langle \Omega_0 \rangle
 +
\langle \Omega_{\rm pair} \rangle, $ where the first term (also
calculated using the replica trick) corresponds to the thermodynamic
potential of unbound fermions $ \langle \Omega_0 \rangle = - T
\sum_{k} \ln \left[ G_{0}^{-1} (k) / T \right]$ and
\begin{equation}
\label{eqn:fermion-green-function}
G_{0}^{-1} (k) = -{\mathrm i} k_m
+ E_{0} (k)
\end{equation}
is the fermion Green's function for unbound fermions with
four-momentum $k = ({\bf k}, {\mathrm i} k_m)$ in the presence of
the disorder potential. Here, $ E_{0} (k) = k^2/2m - \mu +
\Sigma_{0} (k), $ where $ \Sigma_{0} (k) $ is the self-energy for
the unbound fermions in the presence of the disorder potential. This
self-energy can be written as the expansion $ \Sigma_{0} = \kappa
{\partial \Sigma_{0}}/{\partial \kappa} + \frac12 \kappa^2
{\partial^2 \Sigma_{0}}/{\partial \kappa^2} + {\cal O}(\kappa^3)$.
The number of unbound fermions is then given by $N_0 = - \partial
\langle \Omega_0 \rangle/\partial \mu$, and the full number equation
is $ N = -\partial \langle \Omega \rangle /
\partial \mu,
$
where
\begin{equation}
\label{eqn:number-disorder} N (\kappa) = N_0 (\kappa) + N_{\rm pair}
(\kappa)
\end{equation}
contains the contributions
$ N_0 (\kappa) = T \sum_{k} G_{0} (k) {\partial E_{0}(k)}/{\partial
\mu} $ 
for unpaired fermions, and
$ N_{\rm pair} (\kappa) = -T \sum_{q} G_{\rm pair} (q) {\partial
E_{\rm pair} (q)}/{\partial \mu} $ 
for paired fermions.

\begin{figure} [htb]
\psfrag{x}{\LARGE $1/(k_F a_s)$} \psfrag{y}{\LARGE $\eta$}
\psfrag{z}{\LARGE $T_c/\epsilon_F$} \centerline{\scalebox{0.66}
{\includegraphics{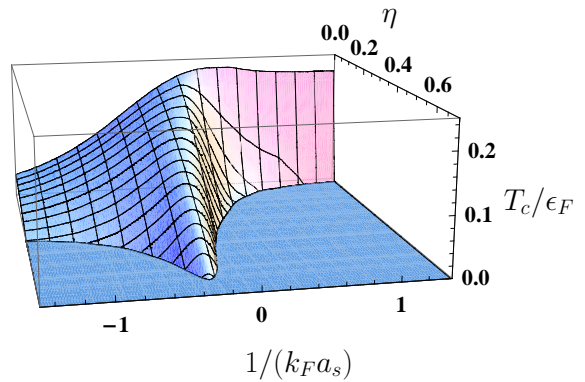} } } \caption{ \label{fig:six}
Critical temperature $T_c/\epsilon_F$ as a function of dimensionless
disorder $\eta$ and interaction $1/(k_F a_s)$ parameters. }
\end{figure}

In our approximation, the solution to
Eqs.~(\ref{eqn:order-parameter-disorder})
and~(\ref{eqn:number-disorder}) produces the critical temperature
$T_c$ as a function of both disorder $\eta$ and interaction
parameter $1/(k_F a_s)$, as shown in Figs.~\ref{fig:five}
and~\ref{fig:six}. Our phase diagram for superfluidity is obtained
at finite $T$ and to second-order in $\kappa$ $(\eta)$, in contrast
to $T = 0$ properties~\cite{orso-08,shklovskii-08} (such as
superfluid density or condensate fraction) obtained to linear order
in $\kappa$ $(\eta)$. Although the replica technique can be used to
all orders in $\kappa$ ($\eta$), we perform numerical calculations
only to second order in $\kappa$ ($\eta$), because it becomes
impractical to calculate higher order contributions to the
self-energies. Therefore, our critical temperatures can be expressed
as $ {\widetilde T}_c (\eta) = {\widetilde T}_c (0) \left[ 1 -
\alpha \eta - \beta \eta^2 \right], $ where ${\widetilde T}_c (0)$,
$\alpha$ and $\beta$ are functions of $1/(k_F a_s)$ only. In
Fig.~\ref{fig:five}, we show $T_c/\epsilon_F$ versus $\eta$ for
three different values of the interaction parameter $1/(k_F a_s)$
corresponding to the BCS side at $1/(k_F a_s) = -1.2$, to unitarity
at $1/(k_F a_s) = 0$, and to the BEC side at $1/(k_F a_s) = 1.2$.
While in Fig.~\ref{fig:six}, we show a three-dimensional plot of
$T_c/\epsilon_F$ as a function of $1/(k_F a_s)$ and $\eta$. Two
important points can be inferred from Figs.~\ref{fig:five}
and~\ref{fig:six} describing the critical temperature. First,
impurity disorder effects seem to be more detrimental in the BEC
regime, when compared to the BCS regime. Second, it is near
unitarity ($(k_F a_s)^{-1} \approx -0.3$) that the superfluid is
more robust to impurity disorder. We attribute this qualitative
difference to the fact that in the BCS regime phase coherence can
only be destroyed through simultaneous breaking of pairs, which does
not occur since the impurity scattering is elastic, and impurities
can not provide the required energy break the pairs. On the other
hand, in the BEC regime the pairing field corresponds to molecular
bosons, and phase coherence can be more easily destroyed without the
requirement of simultaneously breaking pairs.
%
%
%

Even though numerical results are useful, analytical insight into
the evolution from BCS to BEC superfluidity in the presence of
disorder can be obtained through calculations of the critical
temperature to first order in $\eta$, as discussed next.

\section{Analytical results for weak disorder}
\label{sec:analytical-result}

Analytical results to first order in the dimensionless parameter
$\eta$ can be obtained in both the BCS and the BEC limits, where
simplifications occur and calculations are easier, but yet
non-trivial.

In the BCS limit the interaction parameter is large and negative
($1/(k_F a_s) \ll -1 $), and the number equation is dominated
by the contribution due to unbound fermions $N_0 \gg N_{\rm pair}$.
Given that the order parameter equation described in
Eq.~(\ref{eqn:order-parameter-disorder}) is unchanged, then
any change in the critical temperature
$T_c$ has to come from a change in the chemical potential due
to the presence of the impurities. This change can be computed through
the number equation defined in Eq.~(\ref{eqn:number-disorder})
by including only the dominant contribution $N_0$ due to unbound fermions.

The self-energy to linear order in $\kappa$ is $ \Sigma_{0} = \kappa
V^{-1}\sum_{\bf k} ({\mathrm i} \omega - \xi_{\bf k})^{-1} \approx
\eta S(\mu) \left[ \Lambda_c - {\mathrm i} \pi {\rm sgn} (\omega)
\right], $ where $ S(\mu) = (3/2) \sqrt{\epsilon_F \mu}, $ $
\Lambda_c = {\cal P} \int_0^{\epsilon_d/\epsilon_F} {\mathrm d} y
\sqrt{y}/(1 - y), $ and $\epsilon_d/\epsilon_F = 1/(k_F \ell_d)^2$
takes into account the range $\ell_d$ of the disorder correlator
$K({\bf x} - {\bf x}^\prime)$. Replacing $\Sigma_{0}$ into the
number equation, and performing the integrations over four-momentum
leads to a fermion density $ n = N/V \approx n_0 +  \delta n, $
where $ n_0  = 2 \sum_{\bf k} f (\xi_{\bf k}) \approx C \mu^{3/2} $
and $ \delta n = 3\eta C \Lambda_c \sqrt{\epsilon_F} \mu $ with $ C
= (4/3) V^{-1} {\cal N}(\epsilon_F) \epsilon_F^{-1/2}. $ In the case
of perfect particle-hole symmetry $\delta n$ can be neglected and
the chemical potential is {\it pinned} to the Fermi energy $(\mu =
\epsilon_F)$ leading to unchanged $T_c$ given by
\begin{equation}
T_c = \epsilon_F ({\mathrm e}^{\gamma}/\pi) 8 {\mathrm e}^{-2} \exp
[-\pi/(2k_F|a_s|)].
\end{equation}
This reflects Anderson's theorem, as changes in $T_c$ can occur only
via the disorder dependent shift in the chemical potential. Relaxing
the condition of perfect particle-hole symmetry leads to $ \mu =
\epsilon_F \left[ 1 - D \eta \right] $, where $D = 2 \Lambda_c = 4
(\epsilon_d/\epsilon_F)^{1/2}$. The change in $\mu$ produces a
corresponding change in $ {\widetilde T}_c (\eta) = {\widetilde T}_c
(0) \left[ 1 - \alpha \eta \right], $ where $\alpha = \pi D/(4 \vert
k_F a_s \vert)$.

In the BEC regime $(1/(k_F a_s) \gg 1)$ the number equation is
dominated by $N_{\rm pair} \gg N_0$ given that all fermions are
paired into molecular bosons. To linear order in $\kappa$ the pair
self-energy is $ \Sigma_{\rm pair} (q) = g_* {n_{\rm pair}}/{2} +
\kappa_{\rm pair} A (q)$~\cite{lopatin-02}, where $ A (q) = \Lambda
( {\bf q}_c ) - {2m_*^{3/2}}/{(4\pi)} \sqrt{ \vert \bar\mu_* \vert -
{\mathrm i} q_{\ell} }, $ with $ \Lambda ( {\bf q}_c ) = \sum_{\vert
{\bf q} \vert < \vert {\bf q}_{c} \vert } \left[ m_* / (\vert {\bf
q} \vert^2 V) \right] $ and the renormalized chemical potential is $
\bar \mu_* = \mu_* + g_*n_{\rm pair}/2 + \kappa_{\rm pair} \Lambda
({\bf q}_c), $ with $n_{\rm pair} = N_{\rm pair}/V.$ Noticing that
$\partial E_{\rm pair} / \partial \mu \to -2$ in the BEC limit, and
upon summation over Matsubara frequencies ${\mathrm i} q_{\ell}$ in
Eq.~(\ref{eqn:number-disorder}), we arrive at $n_{\rm pair} = 2
n_{B}$, with $ n_B = \zeta (3/2) (m_{*} T_c/2\pi)^{3/2} +
\kappa_{\rm pair} T_c m_{*}^3/4\pi^2$ when $\bar \mu_{*} = 0$. The
solution of the number equation gives the critical temperature
\begin{equation}
\label{eqn:number-bosons}
T_c
(\kappa) = T_c (0) \left[ 1 - \kappa_{\rm pair} T_c (0)
m_{*}^3/6\pi^2 n_B \right],
\end{equation}
where $T_c (0) = 2\pi \left[ n_B/\zeta (3/2) \right]^{2/3}/m_{*}$ is
the Bose-Einstein condensation temperature for a gas of molecular
bosons, which in terms of $\eta$ leads to $ {\widetilde T}_c (\eta)
= {\widetilde T}_c (0) \left[ 1 - \alpha \eta \right]. $ When the
BEC limiting values $T_c (0) \to 0.218 \epsilon_F$, $m_{*} \to 2 m$
and $n_B \to n_F/2$ are used, the coefficient $\alpha = 12\pi^2
(T_c(0)/\epsilon_F) \approx 25.8$ is large.  It is very important to
emphasize that if temporal fluctuations were neglected in the
self-energy $\Sigma_{\rm pair} (q)$ by setting ${\mathrm i} q_\ell =
0$ then one would have come to the conclusion that the critical
temperature $T_c$ for the paired fermions (molecular bosons) is
essentially unaffected by disorder. However, our calculations show
that disorder affects the phase coherence of the molecular bosons
via the incoherent part of $G_{\rm pair} (q)$ manifested in the
branch cut $\sqrt{ \vert \bar\mu_* \vert - {\mathrm i} q_{\ell} }$
of the self-energy $\Sigma_{\rm pair} (q)$. It is these quantum
(temporal) phase fluctuations, which lead to a strong suppression of
$T_c$ in the BEC limit (in comparison to the BCS regime), where
fermions are largely non-degenerate, particle-hole symmetry is
absent, and Anderson's theorem is not applicable.

Now that we have established analytically and numerically the
critical temperature of a disordered superfluid from the BCS to the
BEC limit as a function of the dimensionless impurity parameter
$\eta$, which measures the disorder correlation energy with respect
to the Fermi energy, we would like to comment briefly on the
relation between $\eta$ and the unbound fermion mean free path
$\ell_{F}$ and the relation between $\eta$ and the pair mean
free path $\ell_{\rm pair}$.

\section{Mean free paths}
\label{sec:mean-free-path}

The concept of mean free path $\ell$ is often used in connection
with the classical idea of the average distance travelled by a
particle between collisions with impurities. For quantum particles,
however it is necessary to formulate a more precise meaning for the
mean free path in terms of the details of the impurity potential.
For instance, consider the transition amplitude $ U({\bf x}, {\bf
y}, t) = \langle {\bf y}\vert \exp(-{\mathrm i} {\hat H}t ) \vert
{\bf x} \rangle $ for a quantum particle to propagate from position
${\bf x}$ to position ${\bf y}$ for a given realization of the
impurity potential contained in the Hamiltonian operator ${\hat H}$.
This amplitude can be thought as the sum of all Feynman paths
connecting the two positions, which implies that the action for each
path depends strongly on the each impurity configuration. Taking the
impurity average of the transition amplitude $\langle U({\bf x},
{\bf y}, t) \rangle_{\rm dis}$ leads to the averaging of random
scattering phases, and the expectation that translational invariance
is restored by the averaging process, followed by a rapidly decaying
amplitude: $\langle U({\bf x}, {\bf y}, t) \rangle_{\rm dis} \sim
\exp ( \vert {\bf x} - {\bf y} \vert/\ell)$. The decay constant
$\ell$ is called the elastic mean free path. Notice that the very
definition of $\ell$ requires an exponential decay of the $\langle
U({\bf x}, {\bf y}, t) \rangle_{\rm dis}$. In the cases where this
is not true, it is more difficult to define the elastic mean free
path and thus to use it as a measure of the strength of the impurity
potential.  It is important to emphasize that the concept is only
useful if indeed one can show that this exponential behavior exists.

In two simple limits, we can relate our dimensionless disorder parameter
$\eta$ to an appropriately defined elastic mean free path $\ell$.
For this purpose, we look at the imaginary time description of the
transition probability.

In the BCS limit, the Fourier transformation of the unbound fermion
Green's function $G_0 (k)$ defined in
Eq.~(\ref{eqn:fermion-green-function}) into real space and imaginary
time leads to $ G_{0} ({\bf x}, {\bf y}, \tau) = G_{0,{\rm no}}
({\bf x}, {\bf y}, \tau) \exp \left[ - \vert {\bf x} - {\bf y}
\vert/\ell_F \right]$, where $\ell_F$ is the elastic mean-free-path
of unbound fermions, and $G_{\rm no}$ is the single particle Green's
function in the absence of disorder. In this case, $\ell_F$ can be
calculated analytically as $k_F \ell_F = 4/(3\pi \eta)$ to leading
order in $\eta$. This result shows that when $\eta$ is small then
$k_F \ell_F$ is large.

In the BEC limit, it becomes more appropriate to look at the paired
fermion mean free path. In this case, we look at the Fourier
transform of $G_{\rm pair} (q)$ defined in
Eq.~(\ref{eqn:replica-green-function}), which becomes $ G_{\rm pair}
({\bf x}, {\bf y}, \tau) = G_{{\rm pair}, {\rm no}} ({\bf x}, {\bf
y}, \tau) \exp \left[ - \vert {\bf x} - {\bf y} \vert/\ell_{\rm
pair} \right] $, where $\ell_{\rm pair}$ is generally a function of
$\tau$, and $G_{{\rm pair}, {\rm no}}$ is the pair Green's function
in the absence of disorder. When only low frequency contributions
are included we have $ \ell_{\rm pair} = \sqrt{4\pi} / \left[ 4
m_*^{5/2} \kappa_{\rm pair} \sqrt{g_* n_{\rm pair}/2}
\right]^{1/2}$, which in units of $k_F^{-1}$ becomes $ k_F \ell_{\rm
pair} = F (1/k_F a_s) / \sqrt{\eta}. $ Here,  $F$ is just a function
of $1/k_F a_s$, which takes the limiting value of $F (1/k_F a_s \to
\infty) \approx 0.26$. Notice the non-analytic behavior of $k_F
\ell_{\rm pair}$ when $\eta \to 0$.

In the BCS and BEC cases the phase boundary between the superfluid
state and the normal state occurs prior to the limit where $k_F
\ell_F = 1$, and $k_F \ell_{\rm pair} = 1$, respectively. However,
the present estimates do not include the effects of Anderson
localization, which may become important at low-temperatures $(T
\approx 0)$ and at larger values of $\eta$.

Having made the connection between the dimensionless disorder parameter
$\eta$ and elastic mean free paths for repulsive impurity potentials,
we would like to make next some comparative remarks between speckle and
impurity potentials.

\section{Differences between speckle and impurity potentials}
\label{sec:difference}

Before concluding, we would like to point out some key differences
between speckle and impurity potentials to emphasize that not all
disorders are equal.

The impurity potential described here is purely repulsive, meaning
that it is non-confining, in such a way that for each configuration
of disorder there are no confined single particle states. Thus, the
impurity potential represents essentially scattering centers which
affect mostly the phases of the wave functions of the particles that
are scattered from them. In particular in the BEC regime,
phase-fluctuations of the pairing field occur more easily than in
the BCS regime, and thus impurity effects are more detrimental
relatively speaking.

%
%

In the case of speckle potentials, the situation is quite different,
as there are many {\it mountains} and {\it valleys} in the disorder
potential landscape, which allow for confinement and the existence
of spatially confined states. While in the BCS regime these
differences between speckle and impurity potential are less dramatic
due to the robustness of Cooper pairs to phase fluctuations, the
situation in the BEC regime is different. In the BEC limit the
speckle disorder potential seen by bound pairs allows Bose-Einstein
condensation into the spatially confined states creating many lakes
(local condensates) near the minima of the disorder potential. As
the average amplitude of the speckle potential increases lakes are
formed and become disconnected beyond a certain threshold. Once the
connections between the local condensates is lost, global phase
coherence through the entire cloud is lost, and superfluidity
disappears leading to an insulating state. Thus, the problem of
disorder due to speckle potentials is more like a percolation
problem, and has a percolation threshold. While the confining
regions of the speckle potential favor Bose-Einstein condensation,
overlap between regions of local condensation determine the global
phase coherence and thus superfluidity in the system. Because of
this qualitative difference that favors local Bose-Einstein
condensation, which is a necessary but not sufficient condition for
Bose superfluidity in three-dimensions, we expect that speckle
potentials have less impact on the critical temperature for
superfluidity (as found in quantum Monte Carlo simulations for
atomic Bose systems in speckle potentials~\cite{giorgini-08}) than
the purely repulsive impurity potentials described here. A more
quantitative comparison between the effects of disorder due to
speckle potentials and repulsive impurity potentials requires
modifications of the current approach in order to include the very
important contributions of confining regions, which are of course
absent in the repulsive impurity potential treated here. In
addition, one has to be careful in comparing the sizes of molecular
bosons (produced via Feshbach resonances) and the characteristic
lengths of speckle potentials to determine what is the effective
disorder potential felt by such molecules. These important
theoretical comparisons between different types of disorder are left
for a future publication, but we urge the experimental community to
explore the differences between these types of potentials.

\section{Conclusions}
\label{sec:conclusion}

%
%
In conclusion, we analyzed the effects of repulsive impurity
disorder potentials on the critical temperature for superfluidity of
ultracold fermions during the evolution from the BCS to the BEC
regime. For $s$-wave superfluids, we showed that weak disorder does
not affect the critical temperature of a BCS superfluid with perfect
particle-hole symmetry in accordance with Anderson's theorem, as the
breaking of fermion pairs and the loss of phase coherence occur at
the same temperature. However in the BEC regime, phase coherence is
more easily destroyed by repulsive impurity disorder without the
need of simultaneously breaking fermion pairs. Thus, in the BEC
regime a more dramatic change in the critical temperature occurs,
when compared to the BCS limit. Finally, we also showed that the
superfluid is more robust to the presence of disorder in the
intermediate region near unitarity.



\acknowledgements{We thank NSF (DMR-0709584) and ARO
(W911NF-09-1-0220) for support, and S. Giorgini and N. Prokof'ev for
discussions.}


\end{document}